# Desain dan Implementasi Sistem *Digital Assistant* Berbasis Google Glass pada Rumah Sakit

Daryl Haris Antoni Junior[*1], Muhammad Nur Pratama[*2], Yusrina Nur Dini[*3], dan Ary Setijadi Prihatmanto[*4]

*Program Studi Teknik Elektro*

*Sekolah Teknik Elektro dan Informatika*

*Institut Teknologi Bandung, Jl. Ganesha No. 10, Bandung 40132, Indonesia*

[1]`daryl_pribadi@yahoo.co.id`
[2]`mnurpratama@outlook.com`
[3]`yusrinand@gmail.com`
[4]`asetijadi@lskk.ee.itb.ac.id`

*Abstrak*—**Demi meningkatkan kinerja pelayanan di rumah sakit, salah satu solusi yang diperlukan adalah sebuah sistem yang dapat mempermudah staf medik fungsional dalam menunaikan tugasnya. Makalah ini membahas mengenai desain dan implementasi sistem *digital assistant* yang dirancang untuk digunakan oleh staf medik fungsional di rumah sakit. Sistem ini diimplementasikan menjadi perangkat lunak bernama DIANE pada perangkat keras Google Glass, PC, ponsel pintar, dan tablet. Perangkat Google Glass dilengkapi dengan dua modul yaitu modul data pasien dengan fitur *face recognition* serta modul komunikasi dengan fitur *live streaming*. Perangkat PC diimplementasikan menggunakan ASP.NET Application yang terdiri dari modul *face training*, modul penambahan data, dan modul pengubahan data. Sedangkan perangkat ponsel pintar dan tablet dilengkapi dengan modul personalia dan modul rekam medis. Perangkat tablet turut dilengkapi dengan modul penulisan resep. Seluruh proses pada keempat perangkat keras tersebut turut membutuhkan web service agar masing-masing modul dapat memanipulasi data yang tersimpan pada database di server. Hasil implementasi mengindikasikan bahwa keempat perangkat lunak telah sesuai dengan rancangan dan telah terhubung dengan database SIRS tingkat rumah sakit pada server melalui web service.**

*Kata Kunci*— ***digital assistant*, google glass, rumah sakit.**

## I. PENDAHULUAN

Dalam era globalisasi saat ini, rumah sakit dituntut untuk meningkatkan kinerja dan daya saing sebagai badan usaha dengan tidak mengurangi misi sosial yang dibawanya. Hal ini berarti bahwa rumah sakit harus menerapkan kebijakan-kebijakan strategis dalam hal peningkatan kinerja dan pelayanan, efisiensi dari dalam (organisasi, manajemen, dan sumber daya manusia) serta harus mampu secara cepat dan tepat dalam pengambilan keputusan agar dapat menjadi organisasi yang responsif, inovatif, efektif, dan efisien.

Berangkat dari masalah tersebut, dibuat sebuah sistem *digital assistant* yang bertujuan untuk mempercepat akses staf medik fungsional sebagai seorang profesional dengan waktu terbatas terhadap informasi seperti layanan kesehatan, rekam medis, dan personalia pada SIRS (Sistem Informasi Rumah Sakit). Peningkatan kecepatan akses staf medik fungsional terhadap informasi pada SIRS tingkat rumah sakit dapat dilakukan melalui penggunaan Google Glass yang merupakan *wearable device* berjenis OHMD (*Optical Head-Mounted Display*). Dengan menggunakan Google Glass, staf medik fungsional dapat mengakses informasi pada SIRS tingkat rumah sakit dan menangani pasien dalam waktu bersamaan.

Selain Google Glass, dalam mendukung peningkatan kecepatan akses staf medik fungsional terhadap informasi pada SIRS tingkat rumah sakit, beragam perangkat elektronik seperti PC, ponsel pintar, dan tablet turut digunakan. Ketiga perangkat elektronik tersebut dibutuhkan untuk menjamin ketersediaan informasi seperti layanan kesehatan, rekam medis, dan personalia pada SIRS tingkat rumah sakit.

## II. STUDI PUSTAKA

### A. GOOGLE GLASS

Google Glass merupakan sebuah perangkat OHMD yang berbentuk seperti kacamata. Perangkat tersebut dikembangkan dengan tujuan untuk mempermudah akses manusia terhadap sebuah komputer. Google Glass mampu memberikan pengalaman pengguna seperti sebuah ponsel pintar. Hal tersebut disebabkan Google Glass menggunakan sistem operasi berbasis Android. Penggunanya dapat mengoperasikan Google Glass melalui sentuhan atau perintah suara.

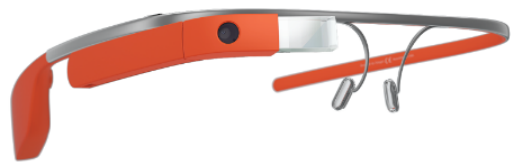

**Gambar 1** Perangkat Keras Google Glass

### B. SISTEM OPERASI ANDROID

Android adalah sistem operasi berbasis Linux yang dirancang untuk beberapa perangkat seluler seperti ponsel pintar, tablet, dan *wearable device* seperti Google Glass. Android adalah sistem operasi *open source* sehingga memungkinkan untuk dilakukannya modifikasi secara bebas. Fitur-fitur pada Android umumnya ditulis dalam bahasa pemograman Java.

Untuk dapat membuat perangkat lunak berbasis Android, umumnya digunakan alat bantu pengembangan berupa perangkat lunak IDE (*Integrated Development Environment*) Android Studio. Di dalam Android Studio terdapat seperangkat alat bantu seperti *debugger*, *library*, *emulator*, dokumentasi, contoh kode, dan tutorial. Dalam pengembangan perangkat lunak berbasis Android pada Google Glass turut dibutuhkan *library* berupa GDK (*Glass Development Kit*).

C. TEKNOLOGI ASP.NET

ASP.NET adalah sebuah *programming framework* berdasarkan *common language runtime* (CLR). Teknologi ASP.NET memungkinkan pengguna untuk menggunakan *server-side* kontrol untuk membangun elemen UI (*user interface*) umum dan mengontrolnya dengan cara pemrograman. Kontrol ini dirancang sehingga mudah untuk membuat sebuah halaman web, serta membuat kode halaman web menjadi lebih ringkas dan efisien. Jika dibandingkan dengan bahasa *interpretative scripting* sebelumnya, ASP.NET memiliki performa yang lebih baik.

Untuk membuat ASP.NET *Web Application* digunakan perangkat lunak IDE (*Integrated Development Environment*) yaitu Microsoft Visual Studio. Microsoft Visual Studio merupakan sebuah perangkat lunak lengkap yang dapat digunakan untuk melakukan pengembangan aplikasi, baik itu aplikasi bisnis, aplikasi personal, ataupun komponen aplikasinya, dalam bentuk aplikasi console, aplikasi Windows, ataupun aplikasi Web. Kompiler yang dimasukkan ke dalam paket Visual Studio antara lain Visual C++, Visual C#, Visual Basic, Visual Basic .NET, Visual J++, Visual J#, dll.

D. WEB SERVICE

Agar sebuah perangkat lunak pada sistem operasi Android dapat memanipulasi data yang tersimpan pada database, diperlukan penghubung antara perangkat lunak dengan database berupa web service. Web service menyediakan autentikasi database sehingga seluruh perangkat lunak yang terhubung dengan web service dapat memanipulasi data yang tersimpan pada database.

Hingga saat ini terdapat dua jenis *communication protocol* yang dapat menghubungkan antara perangkat lunak dengan web service, yaitu SOAP (*Simple Object Access Protocol*) dan REST (*Representational State Transfer*). Keduanya memiliki kelebihan dan kekurangannya masing-masing bergantung pada kebutuhan pengguna.

E. FACE RECOGNITION

*Face Recognition* merupakan sebuah metode pengolahan citra digital untuk mengidentifikasi seseorang menggunakan masukan berupa citra wajah. Saat ini, terdapat beberapa teknik pengolahan citra digital yang cukup terkenal dan dapat digunakan untuk mengidentifikasi identitas pemilik citra wajah seperti *eigen classifier*, *fisher classifier*, dan *local binary pattern histogram classifier*.

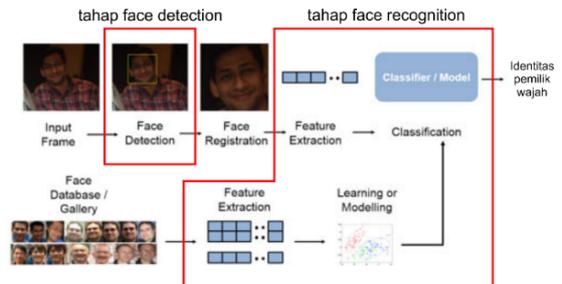

**Gambar 2** Mekanisme Kerja *Face Recognition*

F. FACE TRAINING

*Face training* merupakan sebuah metode pengolahan citra digital untuk menambahkan citra wajah seseorang yang telah berhasil dikenali ke database SIRS tingkat rumah sakit pada server. Citra wajah yang telah disimpan selanjutnya digunakan oleh modul *face recognition* sebagai salah satu referensi dalam proses identifikasi wajah seseorang.

Serupa dengan *face recognition*, *face training* turut menggunakan teknik pengenalan citra wajah seperti *eigen classifier*, *fisher classifier*, atau *local binary pattern histogram classifier*.

G. LIVE STREAMING

*Live Streaming* merupakan sebuah metode untuk dapat mentransmisikan citra dari kamera sebuah perangkat keras secara *realtime* ke server. Citra *realtime* yang diterima oleh server turut dapat diakses dan diunduh oleh pengguna lain.

III. PERANCANGAN PRODUK

A. PERANCANGAN PERANGKAT KERAS

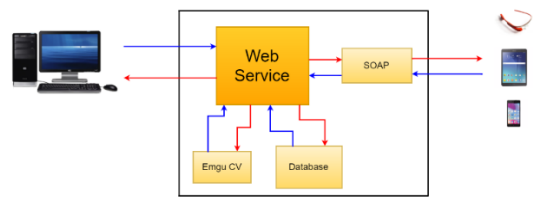

**Gambar 3** Perancangan Perangkat Keras

Sistem ini dibentuk berdasarkan jaringan berarsitektur *client-server*. Sistem ini terdiri dari beberapa perangkat keras seperti Google Glass, PC, ponsel pintar, dan tablet. Keempat perangkat keras merupakan *client* yang berperan sebagai antar muka dengan pengguna yang terhubung dengan server melalui perantara perangkat jaringan. Google Glass merupakan perangkat utama sistem ini sementara tiga perangkat lainnya merupakan perangkat pendukung untuk menjamin ketersediaan informasi yang dibutuhkan oleh perangkat utama. Pada sistem ini, server berperan sebagai pusat proses data dan media penyimpan keseluruhan informasi seperti layanan kesehatan, rekam medis, dan personalia dalam bentuk database SIRS tingkat rumah sakit.

B. PERANGKAT LUNAK PADA GOOGLE GLASS

Berdasarkan masalah-masalah yang muncul kemudian dirancang sebuah perangkat lunak untuk dijalankan pada Google Glass. Spesifikasi perangkat lunak pada Google Glass adalah sebagai berikut:

1. *Face recognition* untuk mengidentifikasi pasien

2. Setelah pasien dikenali, proses akuisisi data pasien berupa biodata dan rekam medis dari database SIRS tingkat rumah sakit pada server dapat dijalankan. Biodata dan rekam medis pasien ditampilkan pada layar Google Glass
3. Komunikasi antarpengguna dengan menggunakan *live streaming*

Dengan rancangan komponen penyusun sebagai berikut.

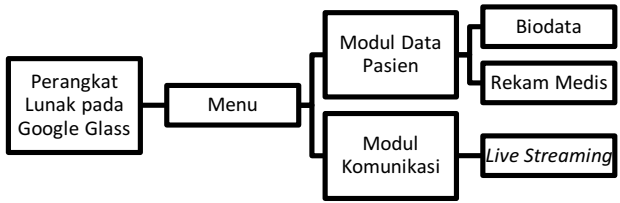

Gambar 4 Komponen Penyusun Perangkat Lunak pada Google Glass

Untuk mendapatkan data pasien berupa biodata dan rekam medis pada modul data pasien, pengguna diminta untuk mengirimkan citra wajah pasien ke web service. Citra wajah tersebut selanjutnya akan menjalani proses *face recognition* di web service agar identitas pasien pemilik citra wajah dapat dikenali.

### C. PERANGKAT LUNAK PADA PC

Berdasarkan kebutuhan data pada Sistem *Digital Assistant* berbasis Google Glass, maka dirancang sebuah perangkat lunak pada PC berbasis ASP.NET *Web Application* yang memiliki spesifikasi sebagai berikut:

1. *Face training* untuk menambah database foto pasien sehingga dapat dikenali melalui Google Glass
2. Penambahan dan pengubahan data personalia baik data pasien, maupun data dokter
3. Penambahan dan pengubahan data rekam medis pasien
4. Penyajian data citra resep

Dengan rancangan komponen penyusun sebagai berikut.

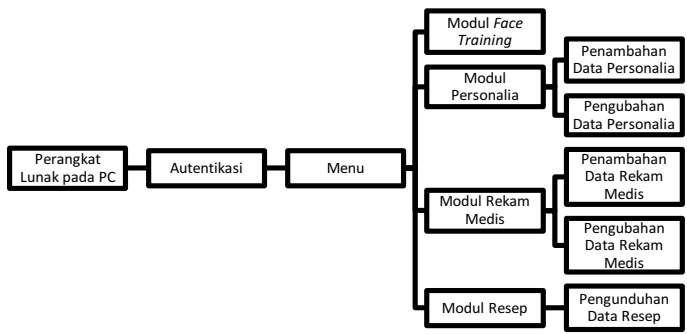

Gambar 5 Komponen Penyusun Perangkat Lunak pada PC

### D. PERANGKAT LUNAK PADA PONSEL DAN TABLET

Karena kebutuhan akses yang informasi yang lebih lengkap dibandingkan dengan yang tertera pada Google Glass, dirancanglah perangkat lunak lainnya pada tablet dan ponsel. Spesifikasi aplikasi pada tablet adalah sebagai berikut:

1. Penulisan resep oleh dokter untuk disimpan pada database SIRS tingkat rumah sakit
2. Akses rekam medis pasien yang lengkap oleh dokter
3. Pengubahan data personalia dokter

Dengan rancangan komponen penyusun sebagai berikut.

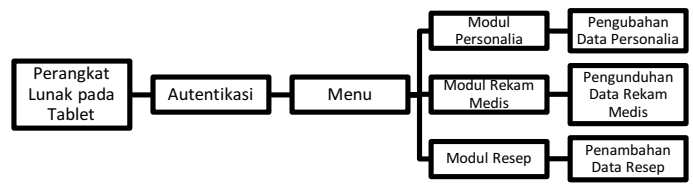

Gambar 6 Komponen Penyusun Perangkat Lunak pada Tablet

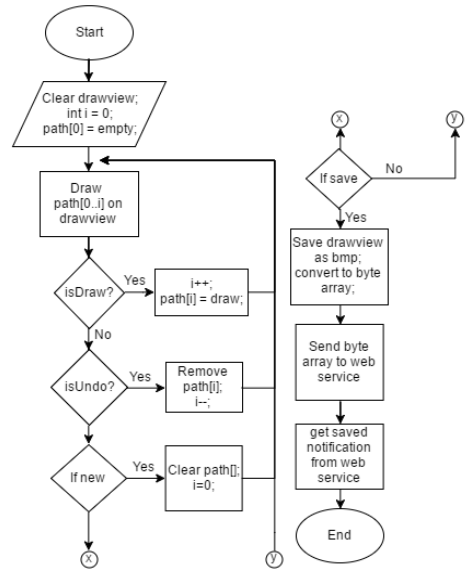

Gambar 7 Diagram Alir Modul Tulis Resep pada perangkat lunak tablet

Sementara itu spesifikasi perangkat lunak pada ponsel adalah sebagai berikut.

1. Rekam Medis: Melalui fitur ini, pasien dapat mengakses rekam medis dirinya secara lebih lengkap
2. Personalia: Pasien dapat mengubah profil diri dan mengirimkan perubahan tersebut ke database SIRS tingkat rumah sakit pada server

Dengan rancangan komponen penyusun sebagai berikut.

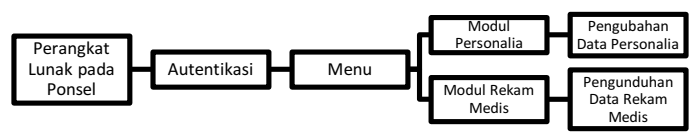

Gambar 8 Komponen Penyusun Perangkat Lunak pada Ponsel

### E. WEB SERVICE PADA SERVER

*Web Service* melakukan proses dan akses terhadap database. Data yang terdapat pada database diolah untuk dikirimkan pada perangkat lunak lainnya sesuai kebutuhan.

Web service yang digunakan pada sistem ini adalah berbasis *communication protocol* SOAP. SOAP memberikan kemudahan dalam perancangan pengembangan *multi platform* dibandingkan *communication protokol* lain misalnya REST.

- Modul Autentikasi

Pada modul autentikasi, *web service* menerima data berupa *email* dan *password*. Data tersebut dibandingkan dengan data pengguna pada database. *Web service* akan mengembalikan hasil verifikasi berupa penanda autentikasi pengguna.

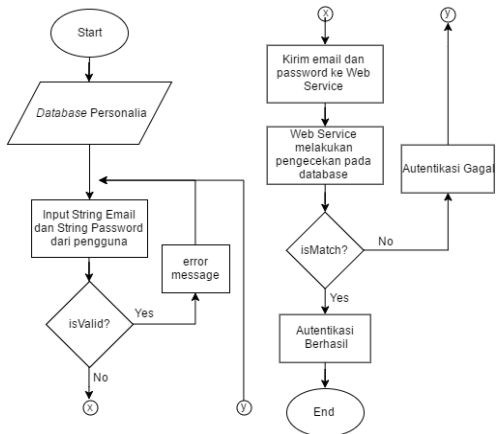

**Gambar 9** Diagram Alir Modul Autentikasi pada *Web Service*

- Modul *Face Recognition* dan *Face Training*

Fitur *face recognition* terdapat di dalam fitur *face training*. Tahapan-tahapan proses modul *face training* yaitu, menerima citra wajah dari halaman web, melakukan *face detection* terhadap citra wajah yang diterima, melakukan *face recognition* ketika wajah terdeteksi, memasukkan data citra wajah ke database ketika *face recognition* tidak menemukan data wajah yang cocok, dan memberikan notifikasi pada halaman web. Tahapan *face detection* dan *face recognition* dibangun dengan menggunakan konstruksi OpenCV pada *platform* Emgu CV.

Perangkat lunak Google Glass menggunakan fitur *face recognition* pada modul data pasien. Modul ini mengambil citra wajah pasien yang kemudian akan dikirimkan pada *web server*. Ketika *face recognition* berhasil mendapatkan citra wajah pasien pada database yang memiliki tingkat kemiripan melebihi *threshold* yang diatur, modul *face recognition* akan mengembalikan data berupa informasi dari pemilik citra wajah tersebut ke modul data pasien pada perangkat lunak di Google Glass

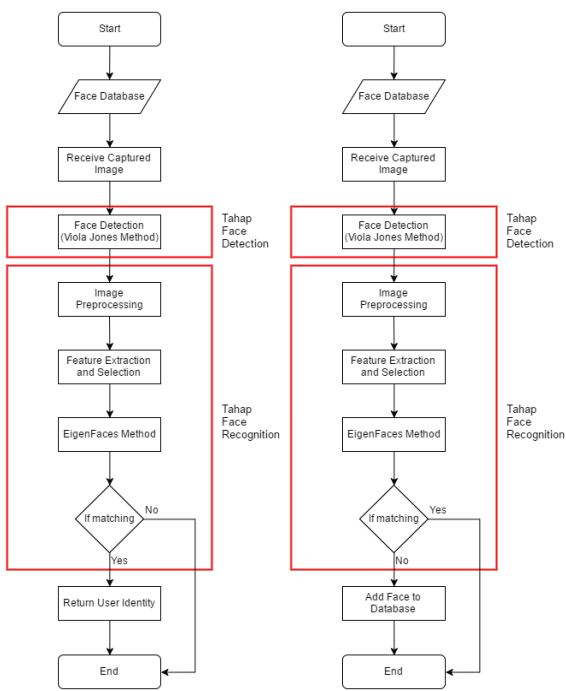

**Gambar 10** Diagram Alir Modul *Face Recognition* (kiri) dan *Face Training* (kanan) pada Web Service di Server

- Modul Pengolahan Data

Modul pengolahan data pada *web service* menerima data dari perangkat lunak berupa ID pasien, ID dokter, data personalia, atau data rekam medis. Data yang diterima dari perangkat lunak akan digunakan untuk menentukan data yang akan diolah pada database baik itu ditambahkan, diperbaharui, maupun hanya diakses untuk melihat data saja.

- Modul Pengunduhan Data Resep

Pada modul pengunduhan resep, *web service* mengirimkan citra berdasarkan data berupa ID dokter yang melakukan akses terhadap web. Hanya citra resep dari dokter yang bersangkutan yang ditampilkan pada web. Pengguna dapat pula melakukan pencarian dengan memberikan data ID pasien pada web sehingga *web service* melakukan pencarian pada database citra resep dengan menggunakan ID dokter yang melakukan akses dan ID pasien yang diberikan.

- Modul Tulis Resep

Melalui modul resep pada perangkat lunak di tablet, pengguna dapat melakukan penambahan data resep. Sebelum dapat melakukan penambahan data resep, pengguna diminta untuk menuliskan resep melalui layar sentuh tablet. Setelah resep berhasil ditulis, didapatkan citra resep. Citra resep selanjutnya dikirimkan ke modul penambahan data pada web service di server.

## IV. HASIL IMPLEMENTASI

### A. PERANGKAT LUNAK PADA GOOGLE GLASS

- Modul Data Pasien

Setelah modul data pasien diaktifkan, maka munculah tampilan berupa hasil pindai citra dari kamera Google Glass. Ketika pengguna memberikan masukan berupa sentuhan pada touchpad, hasil pindai citra berisikan wajah pasien ke modul *face recognition* pada web service di server. Pengiriman citra wajah pasien ke web service dilakukan melalui *communication protocol* SOAP menggunakan jaringan internet..

Jika proses identifikasi citra wajah pasien berhasil dilakukan, maka web service akan mengirimkan kembali data pasien berupa biodata dan rekam medis ke perangkat lunak pada Google Glass.

**Tabel 1** Modul Data Pasien pada Perangkat Lunak di Google Glass

| Tampilan pindai citra dari kamera Google Glass | Pemrosesan citra wajah pasien di modul *face recognition* pada web service di server (setelah *touchpad* Google Glass disentuh) | Tampilan biodata pasien (halaman pertama) |
|---|---|---|
| | | Michael Joshua<br>Male<br>Born on 6/13/1990<br>Patient ID 4 |

| | | Tampilan alergi yang pernah diderita pasien (halaman kedua) 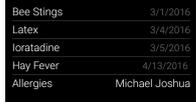 |
| --- | --- | --- |
| | | Tampilan imunisasi yang pernah diberikan ke pasien (halaman ketiga) 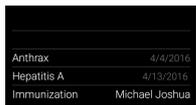 |

- Modul *Live Streaming*

Layanan *live streaming* pada modul komunikasi di perangkat lunak pada Google Glass dibangun berdasarkan konstruksi kickflip. Kickflip menyediakan SDK (*Software Development Kit*) yang dapat digunakan pada pengembangan perangkat lunak berbasis Android. Selain SDK, kickflip juga menyediakan layanan berupa server yang dapat digunakan sebagai media penyimpan citra berupa video hasil *live streaming*.

**Tabel 2** Modul Komunikasi pada Perangkat Lunak di Google Glass

| Citra hasil pindai kamera Google Glass ditampilkan pada layar Google Glass. Selain itu, citra hasil pindai kamera Google Glass turut dikirimkan ke server kickflip melalui jaringan internet agar dapat diunduh oleh pengguna lain melalui website kickflip. |
| --- |
| 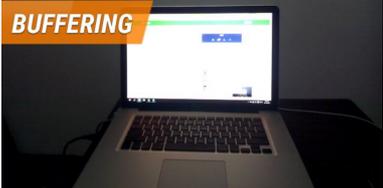 |
| Melalui website kickflip, seorang pengguna lain dapat mengunduh citra hasil pindai kamera Google Glass secara *realtime*. Proses tersebut hanya dapat dilakukan saat modul komunikasi pada perangkat lunak di Google Glass sedang dijalankan. |
| 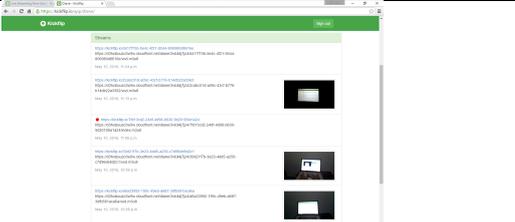 |

B. PERANGKAT LUNAK PADA PC

- Modul Autentikasi

Ketika pengguna melakukan akses pada perangkat lunak pada PC berbasis ASP.NET *Web Application*, halaman web pertama kali yang muncul adalah halaman autentikasi. Autentikasi yang berhasil akan diketahui ketika halaman web berpindah dari halaman autentikasi menjadi halaman menu.

**Tabel 3** Modul Autentikasi pada Perangkat Lunak PC

| Tampilan halaman autentikasi | Tampilan halaman menu ketika autentikasi berhasil |
| --- | --- |
| 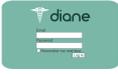 | 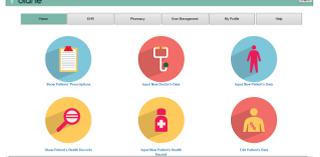 |

- Modul *Face Training*

Setelah data pasien baru dimasukkan, halaman dengan fitur *face training* akan muncul. Halaman ini berisi tampilan citra yang terhubung dengan *webcam* yang terhubung dengan PC. Ketika tombol *capture* ditekan, *webcam* akan mengambil citra yang terlihat dan akan ditampilkan pada halaman tersebut. Pengguna dapat melakukan pengambilan gambar berkali-kali. Ketika tombol *submit* ditekan, citra yang terakhir kali diambil akan dikirimkan ke *web service*.

**Tabel 4** Modul *Face Training* pada Perangkat Lunak PC

| Tampilan halaman *face training* | Tampilan ketika citra diambil oleh *webcam* |
| --- | --- |
| 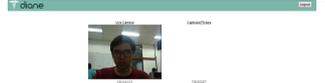 | 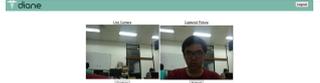 |

Berhasil atau tidaknya *face training* terlihat pada notifikasi yang muncul setelah *submit* citra dilakukan. Notifikasi akan menampilkan "Added" ketika *face training* berhasil dilakukan.

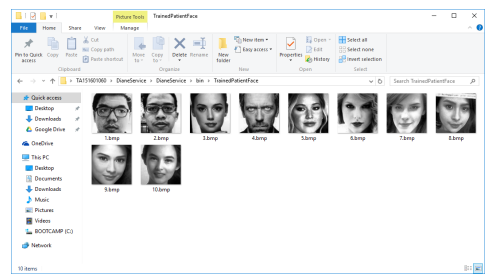

**Gambar 11** Database Citra Wajah pada Server

- Modul Pengolahan Data

Modul pengolahan data terbagi menjadi enam fitur pada halaman web yaitu input data personalia pasien, lihat/update data personalia pasien, input data personalia dokter, lihat/update data personalia dokter, input data rekam medis, dan lihat/update data rekam medis pasien. Fitur lihat/update untuk pasien akan meminta input ID pasien dari pengguna, sedangkan fitur lihat/update untuk dokter menggunakan ID dokter yang sedang melakukan akses terhadap web.

Data yang ditampilkan untuk dilihat dapat diubah oleh pengguna dengan menghapus dan mengisi kembali form yang tertera pada halaman tersebut. Ketika tombol *submit* ditekan, akan mengaktifkan fitur update data sehingga data yang sebelumnya telah tersimpan akan diperbaharui. Data yang ditampilkan pada modul lihat data pun akan menjadi data terbaru.

**Tabel 5** Modul Pengolahan Data pada Perangkat Lunak PC

| Tampilan halaman lihat personalia dokter | Tampilan halaman input data personalia dokter |
|---|---|
| | |
| Tampilan halaman lihat personalia pasien | Tampilan halaman input data personalia pasien |
| | |
| Tampilan halaman lihat rekam medis pasien | Tampilan halaman input data rekam medis pasien |
| | |

- Modul Pengunduhan Resep

Pada halaman modul pengunduhan resep, data citra resep ditampilkan dalam sebuah tabel yang berisi lima baris. Tabel tersebut juga tertera keterangan terkait resep tersebut yaitu nama pasien yang diberikan resep, tanggal diberikan resep, dan status resep tersebut. Citra resep yang berjumlah lebih dari lima akan ditampilkan dengan pengaturan "page" pada halaman web tersebut. Pengguna dapat mencari resep tertentu dengan memberikan ID pasien sehingga dapat dimunculkan resep tertetu yang diberikan pada pasien yang bersangkutan.

**Tabel 6** Modul pengunduhan resep pada Perangkat Lunak PC

| Halaman modul pengunduhan resep |
|---|
| |

C. *PERANGKAT LUNAK PADA PONSEL DAN TABLET*

- Modul Autentikasi

Ketika pengguna melakukan akses pada perangkat lunak pada ponsel dan tablet, halaman yang pertama kali yang muncul adalah halaman autentikasi. Autentikasi yang berhasil akan diketahui ketika halaman berpindah dari halaman autentikasi menjadi halaman utama.

**Tabel 7** Modul Autentikasi pada Perangkat Lunak ponsel dan tablet

| Tampilan halaman autentikasi pada perangkat lunak tablet | Tampilan halaman utama pada perangkat lunak tablet |
|---|---|
| | |
| Tampilan halaman autentikasi pada perangkat lunak ponsel | Tampilan halaman utama pada perangkat lunak ponsel |
| | |

- Modul Pengolahan Data

Modul pengolahan data pada perangkat lunak tablet merupakan modul lihat rekam medis dan modul input/update personalia dokter. Sementara pada perangkat lunak ponsel merupakan modul lihat rekam medis pasien yang melakukan akses dan modul input/update personalia pasien.

**Tabel 8** Modul Autentikasi pada Perangkat Lunak Ponsel dan Tablet

| Tampilan halaman modul rekam medis pada perangkat lunak tablet | Tampilan halaman modul personalia pada perangkat lunak tablet |
|---|---|
| | |
| Tampilan halaman modul rekam medis pada perangkat lunak ponsel | Tampilan halaman modul personalia pada perangkat lunak ponsel |
| | |

- Modul Tulis Resep

Tampilan modul resep merupakan sebuah resep kosong yang belum terisi. Dokter dapat menuliskannya pada layar tersebut kemudian melakukan pengiriman data untuk menyimpan pada database. Fitur lainnya yaitu *undo* unuk menghapus kembali tulisan yang terakhir dan *new* untuk memulai penulisan resep dari awal

| Tampilan halaman tulis resep pada perangkat lunak tablet | Tampilan data citra resep pada database |
|---|---|

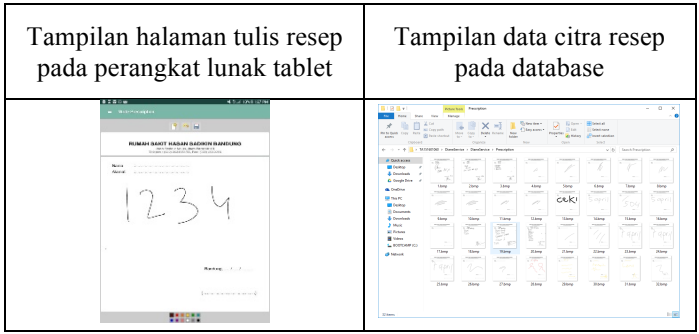

## V. KESIMPULAN

Kesimpulan yang dapat diambil setelah dilakukan implementasi produk adalah sebagai berikut:

1. Sistem Digital Assistant berbasis Google Glass terdiri dari empat perangkat yaitu Google Glass, PC, tablet, dan ponsel pintar.
2. Perangkat lunak Google Glass terdiri dari dua modul yaitu modul data pasien dengan fitur *face recognition* dan modul komunikasi dengan fitur *live streaming*.
3. Perangkat lunak PC terdiri dari fitur *face training*, modul pengolahan data, dan modul pengunduhan data resep.
4. Perangkat lunak tablet terdiri dari modul penulisan resep, modul rekam medis, dan modul personalia.
5. Perangkat lunak ponsel terdiri dari modul rekam medis dan modul personalia, diperuntukkan bagi pasien.
6. Keempat perangkat lunak yang telah dirancang terintegrasi dengan database SIRS tingkat rumah sakit melalui web service.